\documentstyle[prl,aps,psfig]{revtex}
\begin{document}
\twocolumn[\hsize\textwidth\columnwidth\hsize\csname %
@twocolumnfalse\endcsname

\draft
\preprint{CBU-9901}
\title{Reconciling the correlation length for high-spin Heisenberg 
antiferromagnets}
\author{B.B. Beard$^a$, V. Chudnovsky$^b$, and P. Keller-Marxer$^c$ \\
$^a$ Departments of Physics and Mechanical Engineering, Christian Brothers
University, Memphis, TN 38104 \\
$^b$ Department of Physics, Massachusetts Institute of Technology,
Cambridge, MA 02139 \\
$^c$ Institut f\"ur Theoretische Physik, Universit\"at Bern, CH-3012 Bern,
Switzerland \\
}
\date{\today}
\maketitle
\begin{abstract}
We present numerical results for the antiferromagnetic Heisenberg model 
(AFHM) that definitively confirm that chiral perturbation theory, 
corrected for cutoff effects in the AFHM, leads to a correct 
field-theoretical description of the low-temperature behavior of the  
spin correlation length for spins $S\geq 1/2$. With two independent 
quantum Monte Carlo algorithms and a finite-size-scaling technique, we 
explore correlation lengths up to $\xi \approx 10^5$ lattice spacings 
$a$ for spins $S=1$ and $5/2$. We show how the recent prediction of cutoff 
effects by P. Hasenfratz is approached for moderate $\xi/a={\cal O}(100)$, 
and smoothly connects with other approaches to modeling the AFHM at smaller 
correlation lengths.
\end{abstract}
\pacs{75.10.Jm,02.70.Lq,31.15.Kb,71.10.Fd}
]

\narrowtext


In the past decade there has been a resurgence of interest in the 
quantum Heisenberg model. This interest is mainly due to the 
discovery that the undoped, insulating precursors of lamellar high-$T_c$ 
superconducting copper oxides are well-described by the spin $S=1/2$
two-dimensional (2-d) antiferromagnetic Heisenberg model (AFHM) on a 
square lattice. The field theories that describe this model at low
temperatures share the property of asymptotic freedom with the 
theories that describe elementary particles, thus earning a share 
of attention from the high-energy physics community as well.

The low-temperature physics of the AFHM is dominated by magnons.
The magnon interactions are described by the 2-d classical 
continuum $O(3)$ non-linear $\sigma$-model at large correlation lengths. 
This is an extensively studied model in field theory, offering various 
known exact results that can be exploited for the prediction of the 
correlation length $\xi(T)$ in the Heisenberg model at low temperatures. 
The challenge is to find a proper way to connect the parameters of the 
quantum Heisenberg model with the coupling of the $\sigma$-model. Several 
approaches to this problem exist. Chakravarty, Halperin, and Nelson 
\cite{Cha89} used renormalization group arguments to predict the leading 
behavior of $\xi(T)$. Hasenfratz and Niedermayer \cite{Has91,Has90} 
utilized analytical results for the $\sigma$-model to refine the 
prediction, and they used chiral perturbation theory (CPT) 
\cite{Gass84} for the AFHM \cite{Leu90} to connect the parameters 
of the two models.

Neutron scattering experiments on $S=1/2$ 
antiferromagnets such as $\mbox{Sr}_2\mbox{CuO}_2\mbox{Cl}_2$ generally 
agree with this prediction \cite{Gre94}. However, higher-spin 
antiferromagnets have remained problematic. Widely different 
techniques -- including experiment \cite{Nak95,Lee98,Leh98}, 
high-temperature series expansion \cite{Els95}, quantum Monte Carlo 
simulation for $S=1$ \cite{Har97}, and semi-classical approximation 
\cite{Cuc98} -- all showed large deviations from the field-theoretical
prediction by as much as 75\% for $\xi < 200$ lattice spacings, which is
the regime accessible in experiments on $\mbox{La}_2\mbox{Ni}\mbox{O}_4$ 
($S=1$) and $\mbox{Rb}_2\mbox{Mn}\mbox{F}_4$ ($S=5/2$). 
These results suggest that
a serious discrepancy would persist to really large, macroscopic
correlation lengths for $S>1/2$, which is highly unsatisfactory
on theoretical grounds.

In a recent paper \cite{Has99}, Hasenfratz argues that this discrepancy
is due to cutoff effects in the AFHM, which increase strongly with 
spin $S$. When large, they can not be described anymore with the effective 
approach of CPT. Hasenfratz used spin-wave expansion
to calculate the cutoff effects, and he showed the proper way to incorporate
them into the CPT result. 

In this Letter 
we show via extensive quantum Monte Carlo (QMC) calculations that this 
correction indeed accounts for the severe spin dependence of the correlation 
length. Our data connect the regime of large and moderate correlation
lengths, where ``CPT+cutoff" applies, with the regime of small correlation 
lengths where high-temperature and semi-classical results apply. 
The diverse approaches are thereby reconciled.
We also provide evidence that by $S=5/2$ the residual deviation at small
correlation lengths, possibly due to the missing higher-order terms 
in the analytical calculations, has essentially reached the classical 
$S\to\infty$ limit.


Consider the 2-d quantum Heisenberg model with nearest-neighbor
interaction on an $L \times L$ lattice with lattice spacing $a$ and periodic 
boundary conditions, 
\begin{equation}
\label{AFHhamilton}
H = J \sum_{x,\mu} \vec S_x \cdot \vec S_{x+\hat\mu}\;,
\quad \vec{S}^{\,2}_{x}=S(S+1)\;,
\end{equation}
where $J > 0$ is the antiferromagnetic exchange coupling, $\hat{\mu}$
denotes the two primitive translation vectors of the unit cell, and
$\vec S_x$ is the spin operator at position $x$.


The $O(3)$ symmetry of this Hamilton operator is spontaneously broken 
at zero temperature, and the model exhibits 
long-range antiferromagnetic order in the ground state. As a consequence, 
the model has two massless, relativistic Goldstone bosons (called magnons 
or spin waves). However, the Mermin-Wagner-Coleman theorem 
\cite{Mermin66} rules out Goldstone bosons in two dimensions at nonzero 
temperature. Instead, the AFHM magnons acquire a mass 
$c/\xi(T)$ where $c$ is the spin-wave velocity. 
(We set $\hbar$ and $k_{B}$ to unity.) In fact, the 
$\sigma$-model is known to have a non-perturbatively generated mass gap, 
and the leading exponential behavior of the correlation length in the AFHM 
is a consequence of asymptotic freedom in the $\sigma$-model.

The partition function of the 2-d quantum spin model in 
Eq.~(\ref{AFHhamilton}) can be represented by a path integral
\cite{COH} of a classical model with an additional (``time'')
dimension. The continuous coordinate $x_3$
of this periodic Euclidean-time dimension has extent $c/T$. 
When $T\to 0$, the correlation length grows 
exponentially, and becomes much larger than the length 
scale $c/T$. The system then appears dimensionally reduced to a 
thin slab with two infinite space directions and an 
extent  $c/T\ll\xi(T)$ in the Euclidean-time direction. 
This is just a special regularization of the $\sigma$-model.

Hasenfratz and Niedermayer\cite{Has91} used CPT for the AFHM \cite{Leu90}, 
as well as the exact mass gap \cite{Has90} 
and the 3-loop $\beta$-function \cite{Bre78} of the $\sigma$-model 
to derive the asymptotic prediction for the spin 
correlation length in the AFHM:
\begin{equation}
\label{CH_2N_2}
\xi_{\scriptsize \mbox{CH}_2\mbox{N}_2} 
= \frac{e}{8} \frac{c}{2 \pi \rho_{s}}
\exp \left( \frac{2 \pi \rho_{s}}{T} \right)
\left[ 1 - \frac{T}{4 \pi \rho_{s}}
+ {\cal O} \left( T^2 \right) \right].
\end{equation}
The values of spin stiffness $\rho_{s}$ and spin-wave velocity $c$ are not 
fixed by CPT, but they can be estimated by, for example, spin-wave 
expansion (SWE) \cite{Wei91} or fits from QMC data \cite{Wie94,Bea96}. 
Calculating the ${\cal O}(T^2)$ corrections in CPT introduces new, unknown 
parameters. We call Eq.~(\ref{CH_2N_2}) the $\mbox{CH}_2\mbox{N}_2$ 
formula after its parents Chakravarty, Halperin, Nelson, Hasenfratz, and 
Niedermayer.

The QMC data presented in this Letter
confirm that the discrepancy between $\mbox{CH}_2\mbox{N}_2$ and the 
$S>1/2$ AFHM correlation length indeed is severe, and it persists to 
macroscopic correlation lengths $\xi/a \gg 10^5$. 
This situation is unsatisfactory since a mapping of the AFHM
onto the $\sigma$-model must be valid for smaller
correlation lengths, too, due to the dimensional
reduction described above. In particular, this mapping is valid
{\it beyond} the regime of ``renormalized classical scaling'' 
\cite{RC} near $T=0$ (where Eq.~(\ref{CH_2N_2}) 
unquestionably is valid for any $S$). 
The virulent discrepancy indicates a shortcoming in the technique 
that connects the coupling of the $\sigma$-model with the AFHM 
parameters $\rho_{s}$ and $T$ using CPT for large spin.

In his recent calculation \cite{Has99}, Hasenfratz used bilinear 
spin-wave expansion to modify this connection, taking into account 
cutoff effects in the AFHM. In the present study, we also account 
for a minor refinement of the result by Hasenfratz: a part of the 
quadratic temperature dependence, coming from known terms in the 
spin-wave expansion of the coupling of the $\sigma$-model, is 
incorporated. The resulting $\mbox{CH}_3\mbox{N}_2\mbox{B}$ 
formula is (with $t \equiv T/(2\pi\rho_{s})$)
\begin{eqnarray}
\label{CH_3N_2B}
\xi_{\scriptsize \mbox{CH}_3\mbox{N}_2\mbox{B}} 
= \frac{e}{8} \frac{c}{2 \pi \rho_{s}}
\exp \left( \frac{1}{t} \right)
\exp \left( - C(\gamma) \right) \nonumber \\
\times \left[ 1 - \frac{1}{2}t + \frac{27}{32}t^2
+ {\cal O} \left( T^2 \right) \right]\;.
\end{eqnarray}
The parameter $\gamma \equiv 2JS/T$ brings in
the explicit spin dependence. In Ref.~\cite{Has99}, $\exp(-C(\gamma))$ is 
expressed as an integral of familiar spin-wave quantities over the first 
Brillouin zone. The asymptotic $T\to 0$ ($S$ fixed) behavior
is $C(\gamma\to\infty)\sim \gamma^{-2}$, so the $\mbox{CH}_2\mbox{N}_2$
formula, Eq.~(\ref{CH_2N_2}), is recovered in this limit. 
The effect of the aforementioned refinement is simply to add 
the term $(27/32)t^2$ to the polynomial. (This term is
only a part of the ${\cal O}(T^2)$ correction.)  


In Ref.~\cite{Bea98}, an efficient continuous Euclidean-time QMC algorithm 
\cite{Wie94,Bea96,Eve93} was used to study the $S=1/2$ AFHM
correlation length up to 350,000 lattice spacings. For this purpose,
a finite-size-scaling technique, developed by Caracciolo {\it et al.} 
(``CEFPS'') \cite{Car95} for the $\sigma$-model, was applied to the
AFHM finite-volume $\xi(L)$ data. That study confirmed the 
validity of the ``no cutoff effects'' $\mbox{CH}_2\mbox{N}_2$
formula, Eq.~(\ref{CH_2N_2}), but only for large $\xi/a \approx 10^5$. 
By fitting a na\"{\i}ve quadratic term $\alpha t^2$ in the polynomial factor 
of $\mbox{CH}_2\mbox{N}_2$, good agreement was achieved 
down to $\xi/a \approx 100$ (yielding $\alpha=-0.75(5)$).

For $S>1/2$ we used two independent QMC algorithms. 
One is a higher-spin generalization of the $S=1/2$ continuous-Euclidean-time 
loop-cluster algorithm first described in Ref.~\cite{Bea96}. The 
other one is a ``traditional" discrete-Euclidean-time loop-cluster 
algorithm for arbitrary $S$, based on a method proposed by Kawashima and 
Gubernatis \cite{KG}. Results from these two codes were cross-checked 
and agree for all spins within statistical errors, which gives a 
high degree of confidence to our calculations.

The finite-volume correlation lengths $\xi(L)$ were calculated 
using a second-moment method, similar to Eq.~(4.13) in Ref.~\cite{Car93}. 
Afterwards, the CEFPS finite-size-scaling method was applied. 
This method expresses $\xi(2L)/\xi(L)$ as a universal function
$F(\xi(L)/L)$ -- applicable to all models within
the universality class of the 2-d lattice-regularized 
nearest-neighbour $O(3)$ non-linear $\sigma$-model. 
Iteration of $F(\xi(L)/L)$ yields $\xi\equiv\xi(L\to\infty)$.
Unlike the case for $S=1/2$, we found that 
for $S>1/2$ and our level of precision ($\approx 0.2\%$ for 
$\xi(L)$), it is not necessary to incorporate any correction 
for scaling violations, even for  
lattice sizes as small as $L/a \approx 10$.
The maximum $\xi/a$ generated in our study are approximately
170,000 for $S=1$ and 135,000 for $S=5/2$ \cite{web}. 
Note that it is the finite-size scaling technique that enables 
the estimation of correlation lengths much larger than direct 
measurement allows (cf. Ref.~\cite{Har97}, which achieved a maximum
$\xi/a= 24.94(7)$ at $t=0.18$ on a square lattice with side length
$L/a=200$). 
Details of our algorithms, improved estimators, finite-size-scaling 
technique, and calculation of $\xi(L)$ will be provided in a separate 
paper \cite{imp_est}.

\begin{figure}[thb]
\psfig{figure=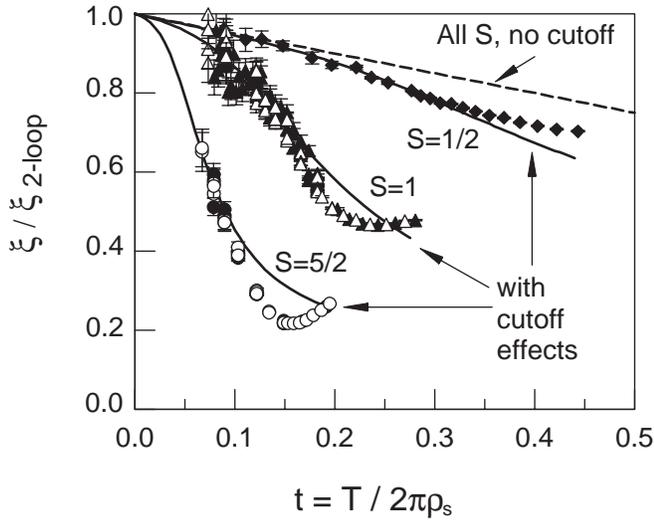,width=8.6cm,clip=}
\caption{
Memphis chart for the AFHM. Correlation length QMC data
and theoretical predictions are normalized by the leading 
(2-loop) prediction $(e/8)(c/2\pi\rho_{s})\exp(1/t)$. Solid lines
are the $\mbox{CH}_3\mbox{N}_2\mbox{B}$ predictions,
Eq.~(\protect\ref{CH_3N_2B}). Dashed line is the spin-independent
``no cutoff effects'' prediction $\mbox{CH}_2\mbox{N}_2$, 
Eq.~(\protect\ref{CH_2N_2}). 
QMC data for $S=1/2$ (diamonds), $S=1$ (triangles), and $S=5/2$ (circles); 
solid (open) symbols are from the continuous (discrete) Euclidean-time algorithm.
The QMC data smoothly merge into $\mbox{CH}_3\mbox{N}_2\mbox{B}$
at low temperatures.
}
\label{memphis}
\end{figure}

Figure~\ref{memphis} shows the QMC data plotted on a Memphis chart, 
where we divide $\xi$ by the 
leading (2-loop) term $(e/8)(c/2\pi\rho_{s})\exp(1/t)$ and plot 
versus $t$. For $S=1/2$, the agreement between QMC and 
$\mbox{CH}_3\mbox{N}_2\mbox{B}$ down to $\xi/a\approx 10$ 
($t\approx 0.3$) is striking. For $S>1/2$
we find that the QMC data smoothly merge into the 
$\mbox{CH}_3\mbox{N}_2\mbox{B}$ predictions at 
$\xi/a > 100$ ($t<0.15$) for $S=1$, and $\xi/a > 500$ 
($t < 0.10$) for $S=5/2$. The agreement in each case degrades 
above some temperature. This is to be expected because
$\mbox{CH}_3\mbox{N}_2\mbox{B}$ leaves out higher-order 
terms from the CPT and spin-wave expansions.
In addition, at high temperatures the field-theoretical requirements 
$\xi \gg c/T$ and $\xi \gg a$ are no longer satisfied,
and predictions such as $\mbox{CH}_3\mbox{N}_2\mbox{B}$ become
meaningless. In our figures we plot the $\mbox{CH}_3\mbox{N}_2\mbox{B}$ 
predictions only for $\xi/a\geq 3$.

Ref.~\cite{Bea98} found that for $S=1/2$, the true value of spin 
stiffness $\rho_{s}$ is about 3\% higher than the value predicted by 
third-order spin-wave theory (SW3). 
We find that comparison of QMC data with $\mbox{CH}_3\mbox{N}_2\mbox{B}$ 
for $S=1/2$ is in agreement with the fitted values of $\rho_{s}=0.1800(5)$ 
and $c=1.657(2)$ found in Ref.~\cite{Bea98}
(we set $J$ and $a$ to unity). That study combined 
the correlation length data fit to $\mbox{CH}_2\mbox{N}_2$, 
Eq.~(\ref{CH_2N_2}), with a fit of finite-volume magnetic 
susceptibilities \cite{Wie94,Bea96} to the predictions of CPT for 
the finite-size and temperature effects in the AFHM \cite{Leu90}. 
High fit precision was achieved by exploiting this combination.
We will present a similar study for the full range of spins 
$S \leq 5/2$ in a separate paper \cite{imp_est}.

For this Letter, we have choosen a different approach which only
involves the correlation length. We demonstrate that for $S>1$ one can
directly rely on the SW3 results \cite{Wei91} to achieve a consistent
connection between the QMC data and
$\mbox{CH}_3\mbox{N}_2\mbox{B}$. For the $S=1$ case we find that the
SW3 predictions $\rho_{s}^{\mbox{\tiny SW3}}=0.869$ and $c^{\mbox{\tiny
SW3}}=3.067$ are nearly correct. Our two-parameter fit gives
$\rho_{s}/\rho_{s}^{\mbox{\tiny SW3}} = 1.005(3)$ and $c/c^{\mbox{\tiny
SW3}} = 0.98(2)$. These ratios correspond to $\rho_{s}=0.8733(23)$ and
$c=3.01(6)$. The fit includes the $\xi/a > 100$ ($t < 0.15$) data in
Figure~\ref{memphis}, and has $\chi^2/\mbox{d.o.f.} = 1.085$ with 58
degrees of freedom (which corresponds to a significance level
$p=30.5\%$). These values of $\rho_{s}$ and $c$ are used in the figures
for $S=1$.

Although the relative deviations from SW3 ($0.5(3)$\% for $\rho_{s}$ and
$-2(2)$\% for $c$) seem small, there is in fact a serious discrepancy. 
Using the SW3 values $(\rho_{s}^{\mbox{\tiny SW3}},c^{\mbox{\tiny SW3}})$ 
in $\mbox{CH}_3\mbox{N}_2\mbox{B}$ and comparing to the QMC $\xi$
gives $\chi^2/60 = 1.39$ (a poor fit, with $p=2.5\%$). Compared to 
the two-parameter fit, SW3 has $\Delta\chi^2=+20.2$ (i.e., outside 
the 99.99\% confidence region). The reason the near-overlap with SW3 is 
deceptive is that the fit parameters $\rho_{s}$ and $c$ are 
strongly anticorrelated (with correlation coefficient $r=-0.977$). 
Interestingly, the major axis of the nearly degenerate error ellipse is
almost orthogonal to the curves of constant
$\theta\equiv\rho_{s}/c^2$. In particular, the $68.3$\% confidence
region for the joint probability distribution of $\rho_{s}$ and $c$
(enclosed by the $\Delta\chi^2=+2.30$ ellipse) intersects the
curve $\rho_{s}/c^2 = \theta^{\mbox{\tiny SW3}}$. In other words, this
$68.3$\% confidence region contains $(\rho_{s},c)$ values which are
consistent with $\theta=\theta^{\mbox{\tiny SW3}}$. We are thus 
motivated to check how close to $\rho_{s}^{\mbox{\tiny SW3}}$ 
and $c^{\mbox{\tiny SW3}}$ the corresponding single-parameter-fit values 
could in fact be.

To do this, we set $\theta_{S=1}=\theta^{\mbox{\tiny SW3}}_{S=1}=0.09238$ 
and performed a fit with the same set of $S=1$ QMC data, with $(\rho_{s},c)$ 
constrained to the one-dimensional parameter subspace 
$\rho_{s}=\theta^{\mbox{\tiny SW3}}c^2$.  
As an upper bound for the error associated with the assumption 
$\theta_{S=1}=\theta^{\mbox{\tiny SW3}}_{S=1}$,
we took from Ref.~\cite{Bea98}
the 4.4\% deviation between $\theta^{\mbox{\tiny SW3}}_{S=1/2}=0.06277$ 
and that study's result $\theta_{S=1/2}=0.06556$. This choice is 
conservative since SWE is an expansion in powers of $1/S$, and  
is expected to become more accurate as spin increases. 
Upon refitting, we found $\rho_{s}/\rho_{s}^{\mbox{\tiny SW3}} =
1.0024(27)$, $c/c^{\mbox{\tiny SW3}} = 1.001(21)$, and $\chi^2/59 =
1.083$ ($p=30.8\%$). These ratios correspond to 
$\rho_{s}=0.8711(24)$ and $c=3.07(6)$. (The difference between 
the one- and two-parameter fits would not be visible in 
Figure~\ref{memphis}.) Note the errors here are dominated by the 
conservative 4.4\% uncertainty in $\theta_{S=1}$; the actual 
errors are bound to be smaller.

For $S=5/2$, we could not identify any deviation from the SW3 values
$\rho_{s}^{\mbox{\tiny SW3}}=5.9444$ and $c^{\mbox{\tiny SW3}}=7.3005$.
We found $\chi^2/14 = 1.194$ ($p=27.2\%$) for the data with $\xi/a > 500$. 
These SW3 values of $\rho_{s}$ and $c$ are used in the figures
for $S=5/2$. 

\begin{figure}[thb]
\psfig{figure=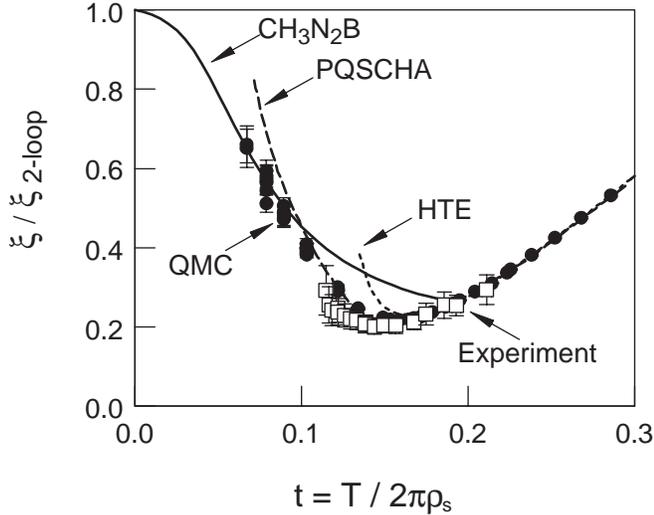,width=8.6cm,clip=}
\caption{
Memphis chart for $S=5/2$, showing $\mbox{CH}_3\mbox{N}_2\mbox{B}$
(solid line); QMC (this work, filled circles);
PQSCHA (dashed line); high-temperature expansion (dotted line);
and experiment (open squares). The QMC data connect the
regimes where the various methods apply.
}
\label{multiplot}
\end{figure}

Figure~\ref{multiplot} shows the situation for $S=5/2$ in more detail.
There is an intermediate regime between $\xi/a \approx 500$ 
($t \approx 0.10$), where $\mbox{CH}_3\mbox{N}_2\mbox{B}$ 
starts to deviate, and $\xi/a \approx 12$ ($t \approx 0.15$), where the 
high-temperature-series expansion (HTE) \cite{Els95} 
starts to fail. Most of the experimental data on 
$\mbox{Rb}_2\mbox{Mn}\mbox{F}_4$\cite{Leh98} fall into this ``gap'',
which exists similarly for $S=1$. At least for
large spin $S=5/2$, this intermediate regime is correctly described by 
the semi-classical approximation known as
the pure-quantum self-consistent harmonic approximation
(PQSCHA) \cite{Cuc98}. The diverse approaches collectively
describe the $S=5/2$ correlation length from extremely small to 
extremely large values.

\begin{figure}[thb]
\psfig{figure=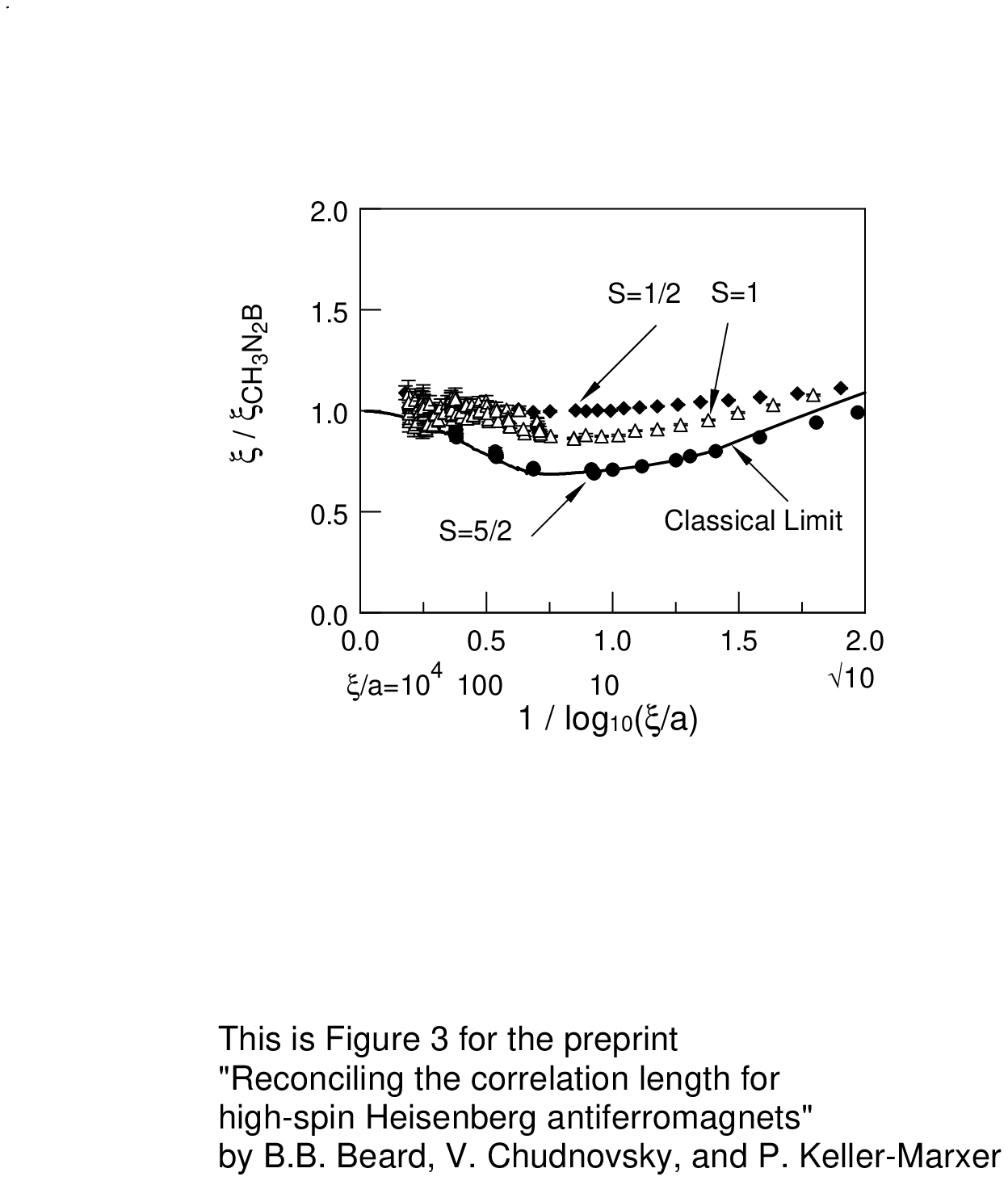,width=8.6cm,clip=}
\caption{
Ratio of correlation length $\xi$ to $\mbox{CH}_3\mbox{N}_2\mbox{B}$ prediction, 
vs. $\xi$, for $S=1/2, 1, 5/2, \infty$. By $S=5/2$ the 
residual deviation essentially reached the classical $S\to\infty$
limit.
}
\label{classical}
\end{figure}

We note that a residual discrepancy between $\mbox{CH}_3\mbox{N}_2\mbox{B}$ 
and numerical data persists in the classical limit $S\to\infty$, where
the AFHM becomes the 2-d lattice-regularized nearest-neighbour $O(3)$ 
non-linear $\sigma$-model. Predictions for this model are available
from analytical calculations \cite{All99} ($\xi/a\geq 10^5$), 
Monte Carlo simulation \cite{Car95,Kim94} ($10\leq\xi/a\leq 10^5$), 
and series expansion \cite{Els95}($\xi/a\leq10$). 
Hasenfratz \cite{Has99} also supplies
the $\gamma \to 0$ form for the correction $\exp(-C(\gamma))$,
which enables the computation of the classical $S\to\infty$ 
limit of Eq.~(\ref{CH_3N_2B}). In Figure~\ref{classical}, we plot 
the ratio of $\xi$ to the $\mbox{CH}_3\mbox{N}_2\mbox{B}$ 
prediction for $S=1/2,1,5/2, 
\mbox{ and } \infty$ versus $1/\log_{10}(\xi)$. 
By $S=5/2$ the discrepancy between the numerical data
and $\mbox{CH}_3\mbox{N}_2\mbox{B}$ has essentially reached the 
classical $S\to\infty$ limit. 
This means that the reasons for the residual discrepancy, including 
finite-order effects of the CPT and spin-wave expansions, are the same 
for the quantum AFHM and the classical $\sigma$-model.

In conclusion, the cutoff correction accounts for the overall spin dependence 
of the correlation length. The spin stiffness and spin-wave velocity approach 
the spin-wave theory predictions rapidly. The diverse
approaches to the AFHM for higher-spin are complementary.

We thank U.-J. Wiese, P. Hasenfratz, F. Niedermayer, and P. Verrucchi
for enlightening discussions. We also thank P. Verrucchi, R. Singh, 
N. Elstner, R.L. Leheny, and R.J. Christianson for use of their data. 
The work of VC was supported in part by the DOE under cooperative research
agreement \#DF-FC02-94ER40818. The work of PKM was supported in part by 
Schweizerischer Natio\-nal\-fonds.


\newpage

\end{document}